\documentclass[10pt,showpacs,preprint]{revtex4-1}
\usepackage{amssymb,amsmath}
\usepackage{graphicx}
\begin{document}
\title{Role of electronic correlations in Ga}
\author{Zhiyong Zhu, Xuhui Wang, and Udo Schwingenschl\"ogl}
\email{udo.schwingenschlogl@kaust.edu.sa}
\affiliation{Physical Sciences \& Engineering Division, KAUST, Thuwal 23955-6900,
Kingdom of Saudi Arabia}

\begin{abstract}
An extended around mean field (AMF) functional for less localized $p$ electrons is
developed to quantify the influence of electronic correlations in $\alpha$-Ga.
Both the local density approximation (LDA) and generalized gradient approximation (GGA)
are known to mispredict the Ga positional parameters. The extended AMF functional
together with an onsite Coulomb interaction of $U_{\rm eff}=1.1$ eV, as obtained from
constraint LDA calculations, reduces the deviations by about 20\%. The symmetry lowering
coming along with the electronic correlations turns out to be in line with the Ga phase diagram.
\end{abstract}

\maketitle

Electronic correlations beyond the local density and
generalized gradient approximation (LDA/GGA) to density functional theory (DFT)
are usually considered to be exclusive to partially filled $d$ and $f$ states
\cite{1,2,3,4}. The localized nature of these orbitals results in a
non-negligible onsite Coulomb interaction $U$ \cite{4a,4b,4c,4d,4e}. 
The importance of correlations in partially filled $p$ orbitals
recently has been pointed out in the context of ``$d^0$ magnetism'', where the
magnetic order arises from the $p$ states instead of the conventional $d$ and $f$
states \cite{5,6}. First-principles calculations have shown that correlations
in open $p$ shells exist not only for first-row elements, like N and O (2$p$), but
also for heavier atoms, like Te (5$p$) \cite{7,7a}. Up to now, the interest in 
$p$ orbital electronic correlations was limited to pure and doped ionic semiconductors.

\begin{figure}[b]
\centering
\includegraphics[width=0.3\textwidth]{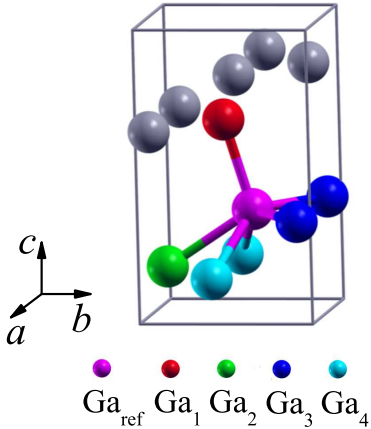}
\caption{\label{fig2} (Color online) Local environment of a reference 
atom Ga$_{\textrm{ref}}$. Ga$_{i}$ ($i=1,2,3,4$) denotes 
the atoms in the $i$-th coordination shell of Ga$_{\textrm{ref}}$. Notice that 
there are two Ga$_{2}$ atoms due to the periodicity along the $b$-axis.}
\end{figure}  

Great interest in $\alpha$-Ga was triggered by the common belief that 
it is the only elemental solid in which metallicity and covalency coexist 
\cite{8}. Showing orthorhombic crystal symmetry with space group 
$Cmca$, $\alpha$-Ga is the most stable Ga phase at ambient conditions 
\cite{9,10}. Besides the lattice constants $a=4.5102$
\AA, $b=4.5161$ \AA, and $c=7.6448$ \AA, the crystal structure is determined by the 
two internal parameters $u=0.0785$ and $v=0.1525$, which specify the Ga atomic positions 
in the unit cell \cite{11}. Every Ga atom (Ga$_{\text{ref}}$ in Fig.\ \ref{fig2}) has one 
nearest neighbor (Ga$_{1}$) and six neighbors in the next three coordination
shells (Ga$_{2}$ to Ga$_{4}$). Partial 
covalency of $\alpha$-Ga, indicated by the short Ga$_{\text{ref}}$-Ga$_{1}$ 
bond, highly anisotropic electronic and thermal conductivities, and a 
steep pseudogap around the Fermi level have been observed in experiments
\cite{12,13,14,15,16} and LDA/GGA calculations \cite{8,11,17}. 
The different bond lengths between Ga$_{\text{ref}}$ and
Ga$_{i}$ ($i=1,2,3,4$) constitute a special local environment, which is 
crucial to the coexistence of metallicity and covalency in $\alpha$-Ga.
As it is reported in Ref.\  \cite{11} and confirmed by our calculations, however, the LDA/GGA
fails miserably to reproduce the experimental Ga positional parameters. It
is our aim to evaluate to which extent electronic correlations account for this deficiency.

The standard way to treat electronic correlations in a DFT calculation is to consider
an onsite parameter $U$ and to embed a Hubbard-like Hamiltonian into the
LDA/GGA Kohn-Sham equations (LDA/GGA+$U$ method) \cite{2}. However, 
usually the onsite $U$ is applied only to the electrons inside the non-overlapping 
muffin-tin spheres. By the neglection of effects from the 
interstitial electrons, any result will depend on the muffin-tin radius
$R_{\text{mt}}$. Major inaccuracies are expected for more delocalized states,
like the 4$p$ electrons of $\alpha$-Ga. As a consequence, an extended 
LDA/GGA+$U$ functional is required which includes the effects of
the interstitial charge. Due to the reduced localization of the $p$ electrons, 
a less strongly correlated state is expected. Hence, the around
mean field (AMF) approximation is chosen for modeling $\alpha$-Ga \cite{2,18}.

Based on the fact that the LDA corresponds to the homogeneous solution of 
the mean-field Hartree-Fock equations with equal occupancy of all sub-orbitals
with the same spin, the AMF energy functional is obtained by supplementing  
the LDA/GGA functional by the additional term \cite{2,4}
\begin{equation}
\label{eq1amf}
E^{\text{AMF}}=-\frac{U_{\text{eff}}}{2}\sum_{\sigma}
Tr(\textbf{n}_{\sigma}-\bar{n}_{\sigma}\textbf{I})^2.
\end{equation}
In this relation $U_{\text{eff}}=U-J$ is the effective interaction, where $U$ and
$J$ denote the onsite Coulomb and exchange interactions, respectively, and
$\textbf{n}_{\sigma}$ the density matrix for spin $\sigma$. In addition,
$\bar{n}_{\sigma}=Tr(\textbf{n}_{\sigma})/(2\ell+1)$ is the average
occupation number of the sub-orbitals with spin $\sigma$ (and orbital
quantum number $\ell$) and \textbf{I} the identity matrix.
The double counting is already corrected in Eq.\ (\ref{eq1amf}).

We assume a linear dependence 
$\textbf{n}_{\sigma}=\boldsymbol{\lambda}_{\sigma}\textbf{n}_{\sigma}^{\text{mt}}$ with
an orbital dependent parameter $\boldsymbol{\lambda}_{\sigma}$.
Because $\textbf{n}_{\sigma}$ and $\textbf{n}_{\sigma}^{\text{mt}}$ can be obtained from
the Mulliken method and the DFT calculations, respectively, 
$\boldsymbol{\lambda}_{\sigma}=\textbf{n}_{\sigma}(\textbf{n}_{\sigma}^{\text{mt}})^{-1}$
can be evaluated. For the Mulliken analysis we will apply the Gaussian code \cite{gaussian}.
In general, $\boldsymbol{\lambda}_{\sigma}$ will depend on the choice of the basis set.
However, we have checked a series of basis sets to ensure that the dependence is sufficiently weak.
To fix the dependence on $R_{\text{mt}}$,
we choose $Tr\boldsymbol{\lambda}_{\sigma}/(2l+1)=2.67$ with $R_{\text{mt}}=2.2$ a.u.\
in the following. The (orbital dependent) one-electron potential now is given by the
functional derivative of $E^{\text{AMF}}$ with respect to $\textbf{n}_{\sigma}^{\text{mt}}$,
\begin{equation}
\label{eq4amf}
\textbf{V}_{\sigma}^{\text{AMF,\,\text{mt}}}=-U_{\text{eff}}\boldsymbol{\lambda}_{\sigma}
(\textbf{n}_{\sigma}-\bar{n}_{\sigma}\textbf{I}).
\end{equation}

\begin{table}[b]
\caption{\label{table1} Fully optimized structural parameters ($a$, $b/a$, $c/a$, 
$u$ and $v$) of $\alpha$-Ga obtained from standard LDA/GGA calculations. Different XC 
functionals, including (1) LDA-PW92, (2) GGA-SOGGA, (3) GGA-AM05, (4) GGA-PBEsol,
(5) GGA-WC, (6) GGA-PBEalpha, (7) GGA-PBE, (8) GGA-PW91, (9) GGA-BPW91, and (10) 
GGA-RPBE, are used \cite{18}. The experimental values are taken from Ref.\ \cite{11}.}
\begin{ruledtabular}
\begin{tabular}{rcccccc}
        &$a$ ($\text{\AA}$)&$b/a$ &$c/a$ &$u$ &$v$ &$\delta{\text{Vol}}$ (\%)\\
 \hline
  Exp & 4.5102 & 1.695 & 1.0013 & 0.0785 & 0.1525 &  0\\
  (1) & 4.4262 & 1.695 & 1.0014 & 0.0843 & 0.1559 & $-5.48$ \\ 
  (2) & 4.4378 & 1.705 & 1.0074 & 0.0821 & 0.1558 & $-3.59$ \\ 
  (3) & 4.4646 & 1.703 & 1.0064 & 0.0815 & 0.1561 & $-2.03$ \\ 
  (4) & 4.4658 & 1.703 & 1.0064 & 0.0829 & 0.1559 & $-1.95$ \\ 
  (5) & 4.4816 & 1.698 & 1.0033 & 0.0834 & 0.1559 & $-1.51$ \\ 
  (6) & 4.5300 & 1.698 & 1.0033 & 0.0825 & 0.1561 & 1.72 \\ 
  (7) & 4.5687 & 1.698 & 1.0033 & 0.0803 & 0.1563 & 4.34 \\ 
  (8) & 4.5743 & 1.695 & 1.0013 & 0.0805 & 0.1564 & 4.32 \\ 
  (9) & 4.5839 & 1.695 & 1.0013 & 0.0826 & 0.1562 & 4.97 \\ 
  (10) & 4.6344 & 1.695 & 1.0013 & 0.0808 & 0.1566 & 8.48 \\ 
\end{tabular}
\end{ruledtabular}
\end{table}

We have implemented the extended AMF approach, given by Eqs.\ (\ref{eq1amf}) and (\ref{eq4amf}),
for the WIEN2k code \cite{18}. Like the original, 
the modified one-electron potential is applied to this part of a
sub-orbital that is inside the muffin-tin sphere. Yet, there are two
improvements: First, the potential exerted by a particular sub-orbital 
arises from the Coulomb repulsion of electrons both inside and outside 
the muffin-tin sphere, $n_{m,\sigma}$, instead of only $n_{m,\sigma}^{\text{mt}}$. 
Second, the parameter $\lambda_{m,\sigma}$ mimicks the energetical effects which
would be obtained if both parts of the sub-orbital (inside
and outside the muffin-tin sphere) were subject to the potential. I.e., 
the extended AMF functional includes effects of charge not only inside the
muffin-tin spheres but also in the interstitial region. Thus, it
eliminates the dependence of the results on $R_{\text{mt}}$,
making it suitable to study electronic correlations in less localized $p$ orbitals.

Using standard LDA/GGA, we have fully optimized the structure of $\alpha$-Ga
against all 5 structural parameters, applying Pulay corrections
to the forces. We use in all calculations $R_{\text{mt}}=2.2$ a.u., $R_{\text{mt}}K_{\text{max}}=8.5$,
and $l_{\text{max}}=10$ as well as the same $23\times23\times14$ $k$-mesh.
The validity of the LDA/GGA is tested for ten exchange-correlation
(XC) functionals, see Table \ref{table1}. In order to provide a quantitative 
comparison between the experimental findings and our results for the unit cell volume (Vol) we study
$\delta{\text{Vol}}=(\text{Vol}-\text{Vol}_{\text{Exp}})/\text{Vol}_{\text{Exp}}$.
The $a$ lattice parameter depends strongly on the
choice of the XC functional, while both ratios $b/a$ and $c/a$ are almost constant.
As a consequence, $\delta{\text{Vol}}$ varies within a wide range from $-5.48\%$ to $8.48\%$. 
The LDA-PW92 functional yields the smallest value of $a$ and thus the smallest volume,
which reflects the common insight that the LDA, in contrast to the GGA, underestimates the
volume \cite{11,20}. The overestimation of $u$ and
$v$ in all our data indicates a systematic failure of the LDA/GGA to describe $\alpha$-Ga.
This may be due to the influence of electronic correlations, which we probe in the following.

\begin{table}[b]
\caption{\label{table2} Experimental \cite{11,12} and calculated positional parameters
and transport anisotropies.}
\begin{ruledtabular}
\begin{tabular}{cccc}
    & Exp & $U_{\rm eff}=0$ eV & $U_{\rm eff}=1.1$ eV \\\hline
$u$ & 0.0785  & 0.0835 & 0.0824 \\
$v$ & 0.1525  & 0.1560 & 0.1556 \\
$\sigma_{c}$:$\sigma_{b}$:$\sigma_{a}$  & 1\,:\,3.1\,:\,6.7 & 1\,:\,1.8\,:\,4.9 & 1\,:\,2.0\,:\,5.3\\ 
$\kappa_{c}$:$\kappa_{b}$:$\kappa_{a}$  & 1\,:\,3.0\,:\,6.6 & 1\,:\,2.0\,:\,4.4 & 1\,:\,1.8\,:\,4.7\\ 
\end{tabular}
\end{ruledtabular}
\end{table}

\begin{figure}
\centering
\includegraphics[width=0.48\textwidth]{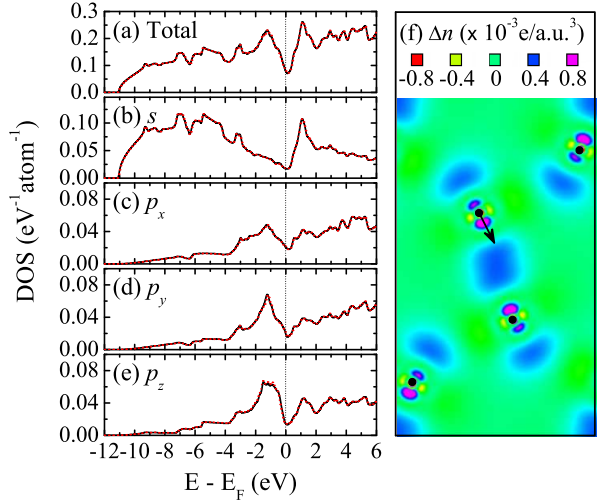}
\caption{\label{fig3} (Color online) (a)-(e) Total and partial DOS obtained
for the GGA-WC (black solid line) and extended AMF (red dotted line)
methods. (f) Charge density difference $\Delta{n}$ in the (100) plane
at $x=0.5$. The positions of the Ga atoms are indicated by black dots.}
\end{figure}

We apply our extended AMF functional to the outer $4p$ orbitals to optimize the crystal
structure, choosing the GGA-WC XC functional for which $\delta{\text{Vol}}$ is minimal.
The computational details are the same as in the LDA/GGA calculations. To
establish the strength of the onsite Coulomb interaction, we employ the constraint
LDA method \cite{19}, which leads to $U_{\rm eff}=1.1$ eV. With this value and the
experimental lattice constants, $u$ and $v$ are optimized in the extended AMF scheme. 
The anisotropies of the electronic conductivity ($\sigma_{c}$:$\sigma_{b}$:$\sigma_{a}$) 
and thermal conductivity ($\kappa_{c}$:$\kappa_{b}$:$\kappa_{a}$)
are obtained \cite{21}, see Table \ref{table2}. The extended AMF
calculations lead to a qualitative improvement of both the structural and 
transport properties. Most remarkably, $u$ is improved by about 20\% with respect to
the experimental value. The extended AMF calculations hence predicts a shorter 
Ga$_{\text{ref}}$-Ga$_{1}$ bond and a slightly higher anisotropy of the electronic
and thermal conductivities. I.e., the partial covalency of $\alpha$-Ga, 
which is underestimated by the GGA-WC, is improved. Still, the total and partial
$s$, $p_{x}$, $p_{y}$, and $p_{z}$ densities of states (DOS) in Figs.\ \ref{fig3}(a)-(e)
reveal only little changes.

On the other hand, an enhanced covalency along the $c$-axis due to the onsite
interaction is reflected by the charge density. In Fig.\ \ref{fig3}(f)
we show the charge density difference within the (100) plane, at $x=0.5$, between the
GGA-WC and the extended AMF results: $\Delta{n}=n_{\text{AMF}}-n_{\text{GGA-WC}}$. An
accumulation of charge is found in the region between Ga$_{\text{ref}}$ and
Ga$_{1}$, indicating a stronger Ga$_{\text{ref}}$-Ga$_{1}$ bond and, thus, an enhanced
covalency. The accumulation comes along with charge transfer from the $p_{x}$ and 
$p_{y}$ sub-orbitals to the $p_{z}$ sub-orbital. The $p_{x}$ and $p_{y}$ occupations
decrease from 0.176 and 0.187 to 0.174 and 0.185, respectively, while
the $p_{z}$ occupation grows from 0.204 to 0.208. The principal component of the electric
field gradient, the direction of which is indicated by the arrow in Fig.\ \ref{fig3}(f),
grows from $4.2\cdot10^{21}$ ${\rm V/m^{2}}$ to $5.3\cdot10^{21}$ ${\rm V/m^{2}}$, whereas
the asymmetry parameter decreases from 0.207 to 0.153. The experimental value is
$5.5\cdot10^{21}$ ${\rm V/m^{2}}$ with an asymmetry parameter of 0.179 \cite{physica04}.
The finite $U_{\rm eff}$ of the 4$p$ electrons is related to the anomalous spatial contraction of the
valence orbitals in $\alpha$-Ga, which is a consequence of incomplete screening of the nuclei by
a relatively shallow Ga 3$d$ state \cite{22}. For the same reason, also a
localization of the 4$s$ electrons would be expected. Unlike the 4$p$ electrons, 
however, the $U_{\rm eff}$ has no effect on the 4$s$ electrons in the AMF approximation \cite{2}.

By Eq.\ (\ref{eq1amf}), a symmetry lowering due to electronic correlations is
expected. Because an asymmetric crystal structure gives asymmetric orbital
populations, $p_{\text{asymm}}=\{\sum_{m,\sigma}[(n_{m,\sigma}-
\bar{n}_{\sigma})/\bar{n}_{\sigma}]^2\}^\frac{1}{2}$, the electronic correlation
energy is related to the asymmetry of the structure by the relation 
$E^{\text{AMF}}=-U_{\text{eff}}\bar{n}^2p^2_{\text{asymm}}/2$, where
$\bar{n}=\bar{n}_\uparrow=\bar{n}_\downarrow$ in non-magnetic $\alpha$-Ga.
Electronic correlations thus stabilize a crystal structure with lower symmetry,
due to a smaller $E^{\text{AMF}}$. This picture is confirmed by the successful
prediction of the quadrupolar lattice distortion in the perovskite compound
KCuF$_{3}$ \cite{3}, where the introduction of the $U$ parameter creates
a ``mexican-hat'' shaped energy surface. This shifts the energy minimum towards
a lattice with lower symmetry. The symmetry lowering due to electronic
correlations is also reflected by the orbital polarization of the extended AMF 
potential functional in Eq.\ (\ref{eq4amf}). If $n_{m,\sigma}$ is larger (smaller) 
than $\bar{n}_{\sigma}$, it further increases (decreases) 
due to a smaller (larger) $V_{m,\sigma}^{\text{AMF,\text{mt}}}$. The orbital 
polarization manifests in the aforementioned occupation numbers of $\alpha$-Ga.

\begin{figure}[t]
\centering
\includegraphics[width=0.42\textwidth]{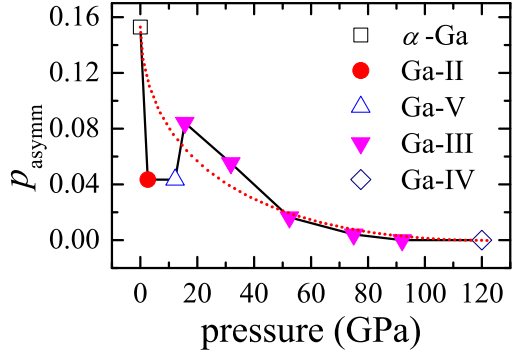}
\caption{\label{fig4} (Color online) Relation between asymmetry, characterized
by $p_{\text{asymm}}$, and stability, characterized by the pressure at which a
phase occurs, for various Ga crystalline phases. The data are taken from
Refs.\ \cite{9,10,23,24}. The red dotted line is a guide to the eye.}
\end{figure}

If electronic correlations are present in $\alpha$-Ga, the
symmetry effect should be seen in the Ga phase diagram.
The Ga-IV phase, which has a face-centered cubic structure, 
occurs only at a pressure above 120 GPa \cite{23}. In contrast, 
the $\alpha$-Ga phase, which has an orthorhombic structure with the lowest
symmetry, is the most stable phase at ambient conditions \cite{9,10}. 
For quantitative description, we use the pressure at which a 
phase occurs as measure of its stability, i.e., higher/lower pressure
corresponds to lower/higher stability. In Fig.\ \ref{fig4} we show
$p_{\text{asymm}}$ as a function of pressure. The results confirm that
structures with lower symmetry give rise to more stable phases.

In conclusion, we have extended the AMF functional to describe correlation
effects of less localized $p$ electrons. The partial covalency of $\alpha$-Ga,
which is underestimated in LDA/GGA calculations, is improved by our approach.
This shows that finite electronic correlations are present in
the partially filled $4p$ orbitals. They are reflected by the crystal
symmetry of the different phases in the Ga phase diagram.

\end{document}